\documentclass[11pt]{article}
\usepackage[dvips]{graphicx}  

\begin{document}

\title{{\bf Novel Gravity Probe B Frame-Dragging Effect}}

\author{{ \bf Reginald T. Cahill}\\
School of Chemistry, Physics and Earth Sciences\\
 Flinders University \\
 GPO Box 2100, Adelaide 5001, Australia \\
 Reg.Cahill@flinders.edu.au 
}

\setcounter{page}{1}

\date{}

\maketitle

\begin{center} arXiv:physics/0406121  June 25, 2004 \end{center}

\begin{center}{\bf Abstract}   \end{center}
The Gravity Probe B (GP-B) satellite experiment will measure the precession of on-board gyroscopes to extraordinary
accuracy.  Such precessions are predicted by General Relativity (GR), and one component of this precession is 
the `frame-dragging'  or Lense-Thirring effect, which is caused by the rotation of the earth.   A new theory of
gravity, which passes the same extant tests of GR, predicts, however,  a second and much larger `frame-dragging'
 precession.  The magnitude and signature of this larger effect is given for comparison with the
 GP-B data.

\section{Introduction}
The Gravity Probe B (GP-B) satellite experiment was launched in April 2004. It has the
 capacity to measure the precession of four on-board gyroscopes to unprecedented
accuracy \cite{Schiff,PEV,Everitt,GPB}.  Such a precession is predicted by the Einstein
theory of gravity, General Relativity (GR), with
 two components (i) a geodetic precession, and (ii) a `frame-dragging' precession
known as the Lense-Thirring effect.  The latter is particularly interesting effect
 induced by the rotation of the earth, and described in GR in terms of a
`gravitomagnetic' field.  According to GR this smaller effect will give
a precession of   0.042 arcsec per year for the GP-B gyroscopes.  However a recently developed  theory  gives
a different account of gravity. While agreeing with GR for all the standard tests of GR
this theory gives a dynamical account of the so-called `dark matter' effect in spiral
galaxies. It also successfully predicts the masses of the black holes found in the
globular clusters M15 and G1.  Here we show that GR and the new theory make very
different predictions for the `frame-dragging' effect, and so the GP-B experiment will
be able to decisively test both theories. While predicting the same earth-rotation
induced precession, the new theory has an additional much larger `frame-dragging'
effect caused by the observed translational motion of the earth. As well the new
theory explains the `frame-dragging' effect in terms of  vorticity in a `substratum
flow'.  Herein the magnitude and signature of this new component of the  gyroscope
precession is predicted for comparison with data from GP-B when it becomes available.

\section{Theories of Gravity}
 The Newtonian `inverse square law' for gravity,
\begin{equation}
F=\frac{Gm_1m_2}{r^2},
\label{eqn:Newton}\end{equation}   was based on Kepler's laws for the motion of the planets.  Newton  formulated
gravity in terms of  the gravitational acceleration vector field
${\bf g}({\bf r},t)$, and in differential form 
\begin{equation}\label{eqn:NG}
\nabla.{\bf g}=-4\pi G\rho,
\end{equation}
where $\rho({\bf r},t)$ is the matter density.  However there is an alternative formulation \cite{NovaDM} in terms of
 a vector `flow' field ${\bf v}({\bf r},t)$ determined by
\begin{equation}
\frac{\partial }{\partial t}(\nabla.{\bf v})+\nabla.(({\bf
v}.{\bf \nabla}){\bf v})=-4\pi G\rho,
\label{eqn:CG1}\end{equation}
with ${\bf g}$ now given by the Euler `fluid' acceleration
\begin{equation}{\bf g}=\displaystyle{\frac{\partial {\bf v}}{\partial
t}}+({\bf v}.{\bf
\nabla}){\bf v}=\displaystyle{\frac{d{\bf v}}{dt}}.
\label{eqn:CG2}\end{equation}
Trivially this ${\bf g}$ also satisfies (\ref{eqn:NG}).
External to a spherical mass $M$ of radius $R$ a velocity field solution of (\ref{eqn:CG1}) is 
\begin{equation}
{\bf v}({\bf r})=-\sqrt{\frac{2GM}{r}}\hat{\bf r},  \mbox{\ \ }r>R,
\label{eqn:vfield}\end{equation}
which gives from (\ref{eqn:CG2}) the usual inverse square law ${\bf g}$ field
\begin{equation}
{\bf g}({\bf r})=-\frac{GM}{r^2}\hat{\bf r}, \mbox{\ \ }r>R.
\label{eqn:InverseSqLaw}\end{equation} 
\index{inverse square law}
However the flow equation (\ref{eqn:CG1}) is not uniquely determined by Kepler's laws because  
\begin{equation}
\frac{\partial }{\partial t}(\nabla.{\bf v})+\nabla.(({\bf
v}.{\bf \nabla}){\bf v})+C({\bf v})=-4\pi G\rho,
\label{eqn:CG3}\end{equation}
where
\begin{equation}
C({\bf v})=\displaystyle{\frac{\alpha}{8}}((tr D)^2-tr(D^2)),
\label{eqn:Cdefn1}\end{equation} and
\begin{equation}
D_{ij}=\frac{1}{2}\left(\frac{\partial v_i}{\partial x_j}+\frac{\partial v_j}{\partial x_i}\right),
\label{eqn:Ddefn1}\end{equation}
also has the same external solution (\ref{eqn:vfield}),  because $C({\bf v})=0$ for the flow in 
(\ref{eqn:vfield}). So the presence of the  $C({\bf v})$ would not have manifested in the special case
of planets in orbit about the massive central sun.
Here $\alpha$ is a   dimensionless constant - a new gravitational constant, in addition to usual
the Newtonian gravitational constant $G$. However inside a spherical mass we find \cite{NovaDM} that 
$C({\bf v})\neq 0$, and using the Greenland  borehole $g$ anomaly data \cite{Greenland} we find that
$\alpha^{-1}=139 \pm  5 $, which gives the fine structure constant $\alpha=e^2\hbar/c \approx 1/137$
to within experimental error. From (\ref{eqn:CG2}) we can write
\begin{equation}\label{eqn:g2}
\nabla.{\bf g}=-4\pi G\rho-4\pi G \rho_{DM},
\end{equation}
where
\begin{equation}
\rho_{DM}({\bf r})=\frac{\alpha}{32\pi G}( (tr D)^2-tr(D^2)),  
\label{eqn:DMdensity}\end{equation} 
which introduces an effective `matter density' representing the flow dynamics associated with the 
$C({\bf v})$ term. In \cite{NovaDM} this dynamical effect is shown to be the `dark matter' effect.
The interpretation of the vector flow field ${\bf v}$ is that it is a manifestation, at the classical
level, of a quantum substratum to space; the flow is a rearrangement of that substratum, and not a flow
{\it through}  space. However  (\ref{eqn:CG3}) needs to be further generalised \cite{NovaDM} to include
vorticity, and also the effect of the motion of matter through this substratum via  
\begin{equation}
{\bf v}_R({\bf r}_0(t),t) ={\bf v}_0(t) - {\bf v}({\bf r}_0(t),t),
\label{eqn:CG$}
\end{equation}
where ${\bf v}_0(t)$ is the velocity of an object, at ${\bf r}_0(t)$, relative to the same frame of reference that defines
the flow field; then ${\bf v}_R$ is the velocity of that matter relative to the substratum. The flow equation
(\ref{eqn:CG3}) is then generalised to, with $d/dt=\partial/\partial t +{\bf v}.\nabla$  the Euler fluid or total
derivative,
\begin{eqnarray}
&&\frac{d D_{ij}}{dt}+ \frac{\delta_{ij}}{3}tr(D^2) + \frac{tr D}{2}
(D_{ij}-\frac{\delta_{ij}}{3}tr D)+\frac{\delta_{ij}}{3}\frac{\alpha}{8}((tr
D)^2 -tr(D^2))\nonumber \\ && +(\Omega D-D\Omega)_{ij}=-4\pi
G\rho(\frac{\delta_{ij}}{3}+\frac{v^i_{R}v^j_{R}}{2c^2}+..),\mbox{ } i,j=1,2,3. 
\label{eqn:CG4a}\end{eqnarray}
\begin{equation}\nabla \times(\nabla\times {\bf v}) =\frac{8\pi G\rho}{c^2}{\bf v}_R,
\label{eqn:CG4b}\end{equation}
\begin{equation}
\Omega_{ij}=\frac{1}{2}(\frac{\partial v_i}{\partial x_j}-\frac{\partial v_j}{\partial
x_i})=-\frac{1}{2}\epsilon_{ijk}\omega_k=-\frac{1}{2}\epsilon_{ijk}(\nabla\times {\bf v})_k,
\label{eqn:BS}\end{equation}
and the vorticity vector field is $\vec{\omega}=\nabla\times {\bf v}$. For zero vorticity and
$v_R\ll c$ (\ref{eqn:CG4a}) reduces to (\ref{eqn:CG3}).   We obtain from
(\ref{eqn:CG4b}) the Biot-Savart form for the vorticity 
\begin{equation}
\vec{\omega}({\bf r},t)
=\frac{2G}{c^2}\int d^3 r^\prime \frac{\rho({\bf r}^\prime,t)}
{|{\bf r}-{\bf r}^\prime|^3}{\bf v}_R({\bf r}^\prime,t)\times({\bf r}-{\bf r}^\prime).
\label{eqn:omega}\end{equation}

The path ${\bf r}_0(t)$ of an object through this flow  is obtained by  extremising  the relativistic
proper time
\begin{equation}
\tau[{\bf r}_0]=\int dt \left(1-\frac{{\bf v}_R^2}{c^2}\right)^{1/2}
\label{eqn:CG5}\end{equation}
giving, as a generalisation of (\ref{eqn:CG2}), the acceleration
\begin{equation}\label{eqn:CG6}
 \frac{d {\bf v}_0}{dt}=-\frac{{\bf
v}_R}{1-\displaystyle{\frac{{\bf v}_R^2}{c^2}}}
\frac{1}{2}\frac{d}{dt}\left(\frac{{\bf v}_R^2}{c^2}\right)
+\left(\displaystyle{\frac{\partial {\bf v}}{\partial t}}+({\bf v}.{\bf \nabla}){\bf
v}\right)+({\bf
\nabla}\times{\bf v})\times{\bf v}_R.
\end{equation}
Formulating gravity in terms of a flow is probably unfamiliar, but General Relativity (GR) permits  an
analogous result for metrics of the Panlev\'{e}-Gullstrand class \cite{PG},
\begin{equation}
d\tau^2=g_{\mu\nu}dx^\mu dx^\nu=dt^2-\frac{1}{c^2}(d{\bf r}-{\bf v}({\bf r},t)dt)^2.
\label{eqn:PGmetric}\end{equation}
The external-Schwarzschild metric belongs to this class  \cite{RGC}, and when expressed in the form of (\ref{eqn:PGmetric}) the ${\bf
v}$ field is identical to (\ref{eqn:vfield}).  Substituting (\ref{eqn:PGmetric}) into the Einstein equations
 \begin{equation}
G_{\mu\nu}\equiv R_{\mu\nu}-\frac{1}{2}Rg_{\mu\nu}=\frac{8\pi G}{c^2} T_{\mu\nu},
\label{eqn:EG}\end{equation}
gives
\begin{eqnarray}\label{eqn:EG1}
G_{00}&=&\sum_{i,j=1,2,3}v_i\mathcal{G}_{ij}
v_j-c^2\sum_{j=1,2,3}\mathcal{G}_{0j}v_j-c^2\sum_{i=1,2,3}v_i\mathcal{G}_{i0}+c^2\mathcal{G}_{00}, 
\nonumber\\ G_{i0}&=&-\sum_{j=1,2,3}\mathcal{G}_{ij}v_j+c^2\mathcal{G}_{i0},   \mbox{ \ \ \ \ } i=1,2,3.
\nonumber\\ G_{ij}&=&\mathcal{G}_{ij},   \mbox{ \ \ \ \ } i,j=1,2,3.
\end{eqnarray}
where the  $\mathcal{G}_{\mu\nu}$ are  given by
\begin{eqnarray}\label{eqn:EG2}
\mathcal{G}_{00}&=&\frac{1}{2}((trD)^2-tr(D^2)), \nonumber\\
\mathcal{G}_{i0}&=&\mathcal{G}_{0i}=-\frac{1}{2}(\nabla\times(\nabla\times{\bf v}))_i,   \mbox{ \ \ \ \ }
i=1,2,3.\nonumber\\ 
\mathcal{G}_{ij}&=&
\frac{d}{dt}(D_{ij}-\delta_{ij}trD)+(D_{ij}-\frac{1}{2}\delta_{ij}trD)trD\nonumber\\ & &
-\frac{1}{2}\delta_{ij}tr(D^2)+(\Omega D-D\Omega)_{ij},  \mbox{ \ \ \ \ } i,j=1,2,3.
\end{eqnarray}
and so GR also uses the Euler `fluid' derivative,  and we obtain a set of equations
analogous but not identical to (\ref{eqn:CG4a})-(\ref{eqn:CG4b}). In vacuum, with $T_{\mu\nu}=0$, we find that 
(\ref{eqn:EG2}) demands that
\begin{equation}
((trD)^2-tr(D^2))=0.
\label{eqg:EG3}\end{equation}  
This simply corresponds to the fact that GR does not permit the `dark matter' dynamical  effect, namely that
$\rho_{DM}=0,$ according to (\ref{eqn:DMdensity}). This happens because GR was forced to agree with
Newtonian gravity, in the appropriate limits, and that theory also has no such effect.  The predictions
from (\ref{eqn:CG4a})-(\ref{eqn:CG4b}) and from (\ref{eqn:EG2}) for the Gravity Probe B experiment are
different, and provide an opportunity to test both gravity theories.

\section{`Frame-Dragging' as a Vorticity Effect} 

\begin{figure}[t]
\hspace{35mm}\includegraphics[scale=1.0]{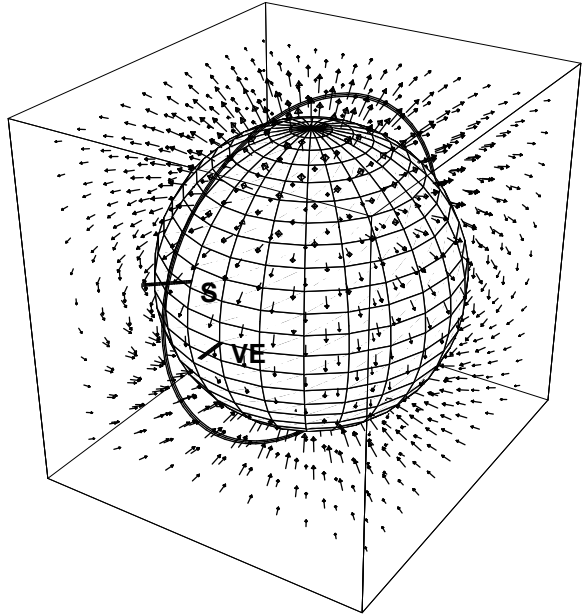}
\caption{\small{ Shows the earth (N is up) and vorticity vector field component $\vec{\omega}$ induced by the rotation of the
earth, as in  (\ref{eqn:rotation}). The polar orbit of the GP-B satellite is shown,    ${\bf S}$ is the gyroscope starting 
spin orientation, directed towards the guide  star IM Pegasi, RA = $22^h $ $53^\prime$ $ 2.26^{\prime\prime}$, Dec = $16^0$ $
50^\prime $ $28.2^{\prime\prime}$, and  ${\bf VE}$ is the vernal equinox.}
\label{fig:Rotation}}\end{figure}

\begin{figure}[t]
\hspace{35mm}\includegraphics[scale=1.0]{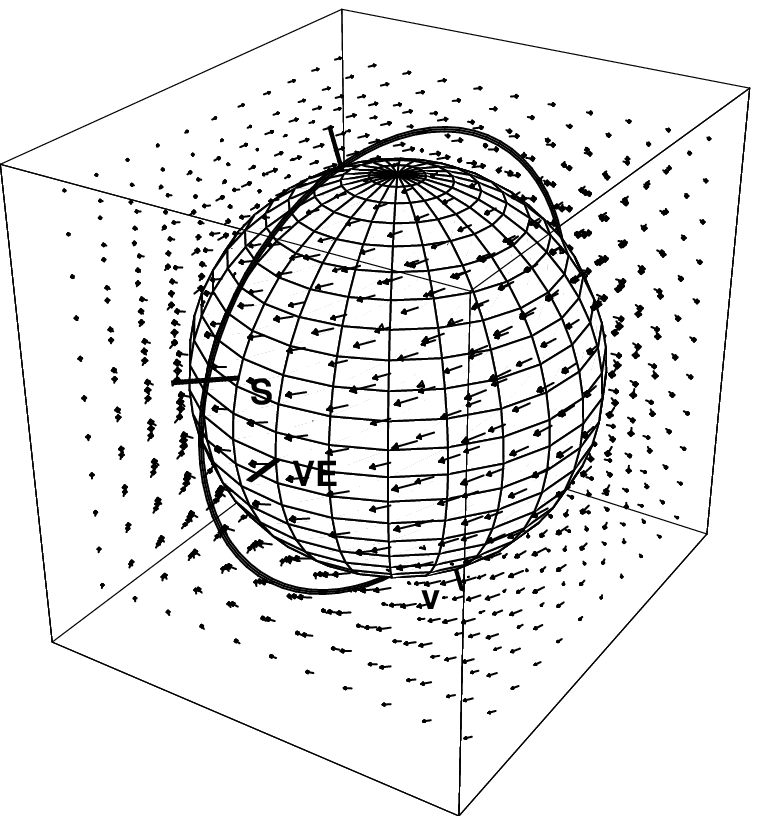}
\caption{\small{ Shows the earth (N is up) and the much larger vorticity vector field component $\vec{\omega}$ induced by the
translation of the earth, as in  (\ref{eqn:AMomega}). The polar orbit of the GP-B satellite is shown, and   ${\bf S}$ is the
gyroscope starting  spin orientation, directed towards the guide  star IM Pegasi,  RA = $22^h $ $53^\prime$ $ 2.26^{\prime\prime}$, Dec
= $16^0$ $ 50^\prime $ $28.2^{\prime\prime}$,   ${\bf VE}$ is the vernal equinox,
and ${\bf V}$ is the direction  $\mbox{RA} = 5.2^{h}$, $\mbox{Dec} = -67^0$ of the translational velocity ${\bf v}_c$.}
\label{fig:Absolute}}\end{figure}

Here we consider one difference between the two theories, namely that associated with the vorticity
part of  (\ref{eqn:CG6}), leading to the `frame-dragging' or Lense-Thirring  effect. In GR the vorticity field
 is known as the `gravitomagnetic' field ${\bf B}=-c\:\vec{\omega}$. In  both GR and  the new
theory the vorticity is given by (\ref{eqn:omega}) but with a key difference: in GR ${\bf v}_R$ is {\it only} the
rotational velocity of the matter in the earth, whereas in (\ref{eqn:CG4a})-(\ref{eqn:CG4b})
${\bf v}_R$ is the vector sum of the  rotational velocity and the translational velocity of the
earth through the substratum.   At least seven experiments have detected this translational velocity; some
were gas-mode Michelson interferometers and others coaxial cable experiments \cite{RGC,AMGE,NovaBook}, and the
translational velocity is now known to be  approximately  430 km/s in the direction RA $ = 5.2^{h}$,
Dec$ = -67^0$. This direction has been known since  the Miller
\cite{Miller} gas-mode interferometer experiment, but the RA was more recently confirmed by the 1991
 DeWitte coaxial cable experiment performed in the Brussels laboratories of Belgacom \cite{AMGE}. 
This flow is related to galactic gravity flow effects  \cite{RGC,AMGE,NovaBook}, and so  is different to that
of the velocity of the earth with respect to the Cosmic Microwave Background (CMB), which is $369$ km/s in the direction
 $\mbox{RA }=11.20^h,\mbox{Dec } =-7.22^0$.

First consider the common but much smaller rotation induced `frame-dragging' or vorticity effect. Then
${\bf v}_R({\bf r})={\bf w}\times{\bf r}$ in (\ref{eqn:omega}), where ${\bf w}$ is the angular
velocity of the earth, giving
\begin{equation}
\vec{\omega}({\bf r})=4\frac{G}{c^2}\frac{3({\bf r}.{\bf L}){\bf r}-r^2{\bf L}}{2 r^5},
\label{eqn:rotation}\end{equation}
where ${\bf L}$ is the \index{angular momentum - earth} angular momentum of the earth, and ${\bf
r}$ is the distance from the centre. This component of the vorticity field is shown in
Fig.\ref{fig:Rotation}.  Vorticity may be detected by observing the precession of the GP-B
gyroscopes.  The vorticity term in 
 (\ref{eqn:CG6}) leads to a torque on the angular momentum ${\bf S}$ of the gyroscope,
\begin{equation}
\vec{\tau}= \int d^3 r \rho({\bf r})\; {\bf r}\times(\vec{\omega}({\bf r}) \times{\bf v}_R({\bf r})),
\label{eqn:torque1}\end{equation}
where $\rho$ is its  density, and where
  ${\bf v}_R$ is used here to describe the rotation of the gyroscope.  Then $d{\bf S}=\vec{\tau}dt$ is the change in
${\bf S}$ over the time interval $dt$. In the above case 
${\bf v}_R({\bf r})={\bf s}\times{\bf r}$, where ${\bf s}$ is the angular velocity of the gyroscope.  
This gives
\begin{equation}
\vec{\tau}=\frac{1}{2}\vec{\omega}\times{\bf S}
\label{eqn:torque2}\end{equation}
and so $\vec{\omega}/2$ is the instantaneous angular velocity of precession of the gyroscope. This corresponds to
the well known fluid result that the vorticity vector is twice the angular velocity vector.   For GP-B the direction
of
${\bf S}$   has been chosen so that this precession is cumulative and, on averaging  over an orbit,
corresponds to some $7.7\times 10^{-6}$ arcsec per orbit, or 0.042 arcsec per year.  GP-B has been superbly
engineered so that measurements to a precision of 0.0005 arcsec are possible. 

\begin{figure}[t]
\hspace{10mm}\includegraphics[scale=1.5]{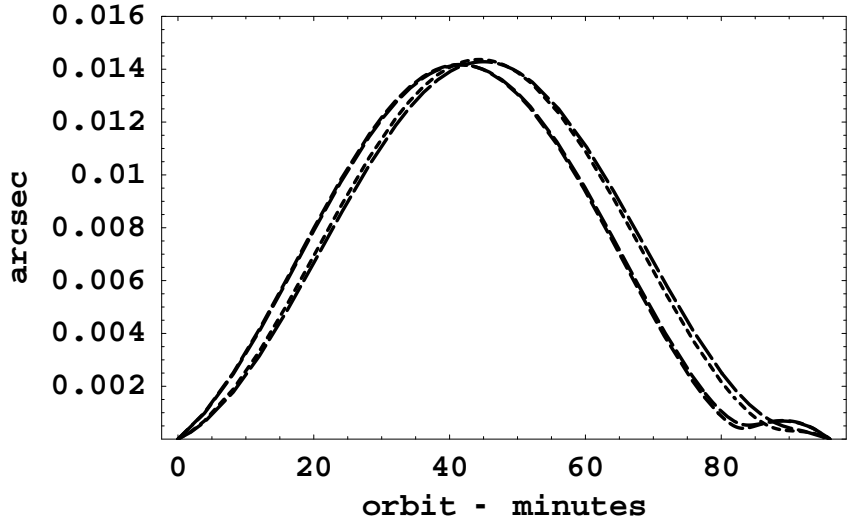}
\caption{\small{ Predicted  variation of the precession angle  $\Delta \Theta=|\Delta {{\bf S}}(t)|/|{\bf S}(0)|$, in
arcsec, over one 97 minute GP-B orbit, from the vorticity induced by the translation of the earth, as given by
(\ref{eqn:precession}). The orbit time begins at location
${\bf S}$. Predictions are for the  months of April, August, September and February, labeled by  increasing dash
length.  The `glitches' near 80 minutes are caused by the angle effects in (\ref{eqn:precession}). These changes arise
from the effects of the changing orbital velocity of the earth about the sun.  The GP-B expected angle measurement
accuracy is 0.0005 arcsec. Novel gravitational waves will  affect these plots. }
\label{fig:Precession}}\end{figure}

However for the unique translation-induced precession if we  use $v_R \approx v_C = 430$ km/s in the
direction  $\mbox{RA} =5.2^{hr}$, $\mbox{Dec} =-67^0$, namely ignoring the effects of the orbital motion of the
earth, the observed flow past the earth towards the sun, and the flow into the earth, and effects of
the gravitational waves, then (\ref{eqn:omega}) gives
\begin{equation}
\vec{\omega}({\bf r})=\frac{2GM}{c^2}\frac{{\bf v}_C\times{\bf r}}{r^3}.
\label{eqn:AMomega}\end{equation}
This much larger component of the vorticity field is shown in Fig.\ref{fig:Absolute}.
The maximum magnitude of the speed of this precession  component is $\omega/2=gv_C/c^2=8 \times10^{-6}$arcsec/s, where here
$g$ is the gravitational acceleration at the altitude of the satellite.   This precession has a different signature: it  is
not cumulative, and is detectable by its variation over each single orbit, as its orbital average is zero, to first
approximation.   Fig.\ref{fig:Precession} shows   $\Delta \Theta=|\Delta {{\bf S}}(t)|/|{\bf S}(0)|$  over 
 one orbit, where, as in general,
\begin{equation}\Delta {{\bf S}}(t) =
\int_0^t dt^\prime \frac{1}{2}\vec{\omega}({\bf r}(t')) \times {\bf S}(t^\prime)
\approx \left(\int_0^t dt^\prime \frac{1}{2}\vec{\omega}({\bf r}(t'))\right) \times
{\bf S}(0).
\label{eqn:precession}\end{equation}  
Here $\Delta {{\bf S}}(t)$ is the integrated change in spin, and where
the approximation arises  because the change in
${\bf S}(t^\prime)$ on the RHS of (\ref{eqn:precession}) is negligible.   The plot in  Fig.\ref{fig:Precession}  shows
this effect to be some 30$\times$ larger than the expected GP-B errors, and so easily detectable, if it exists as
predicted herein.  This precession is about the instantaneous direction of the vorticity $\vec{\omega}({\bf r}((t))$
at the location of the satellite, and so is neither in the plane, as for the geodetic precession, nor perpendicular to
the plane of the orbit, as for the earth-rotation induced vorticity effect. 

Because the yearly orbital rotation  of the earth about the sun slightly effects 
${\bf v}_C$ \cite{AMGE} predictions for four months throughout the  year are shown  in
Fig.\ref{fig:Precession}. Such yearly effects were first seen in the Miller \cite{Miller}
experiment.

The other non-vorticity acceleration terms in  (\ref{eqn:CG6}) also result in torques on the gyroscope, and the magnitude and
signature of the resultant precessions will be reported elsewhere, and are required in the data analysis.

\end{document}